\documentclass[a4paper,pra,showpacs,superscriptaddress,nofootinbib,reprint]{revtex4-1}
\usepackage[utf8]{inputenc}
\usepackage[T1]{fontenc}
\usepackage[UKenglish]{babel}
\usepackage{lmodern}
\usepackage[colorlinks]{hyperref}
\usepackage{graphicx}
\usepackage{amsmath,amssymb,amsthm}
\usepackage{xspace}
\usepackage{mathtools}
\usepackage{dsfont}
\usepackage{lipsum}
             \usepackage{siunitx}
             \usepackage{pgf,tikz}
             \usepackage{pgfplots}

\DeclareMathOperator{\tr}{tr}

\let\originalleft\left
\let\originalright\right
\renewcommand{\left}{\mathopen{}\mathclose\bgroup\originalleft}
\renewcommand{\right}{\aftergroup\egroup\originalright}

\newcommand{\ket}[1]{\left| #1 \right\rangle}

\newcommand{\proj}[1]{\left|#1\middle\rangle\!\middle\langle#1\right|}

\newcommand{\mathand}{\quad\text{and}\quad}

\newcommand{\Hi}{\mathcal{H}}

\newcommand{\id}{\mathds{1}}

\newcommand{\ie}{\emph{i.e.}\@\xspace}

\begin{document}

\title{The simplest causal inequalities and their violation}
\author{Cyril Branciard}
\affiliation{Institut Néel, CNRS and Universit\'e Grenoble Alpes, 38042 Grenoble Cedex 9, France}
\author{Mateus Araújo}
\affiliation{Faculty of Physics, University of Vienna, Boltzmanngasse 5 1090 Vienna, Austria}
\affiliation{Institute for Quantum Optics and Quantum Information (IQOQI), Boltzmangasse 3 1090 Vienna, Austria}
\author{Adrien Feix}
\affiliation{Faculty of Physics, University of Vienna, Boltzmanngasse 5 1090 Vienna, Austria}
\affiliation{Institute for Quantum Optics and Quantum Information (IQOQI), Boltzmangasse 3 1090 Vienna, Austria}
\author{Fabio Costa}
\affiliation{Faculty of Physics, University of Vienna, Boltzmanngasse 5 1090 Vienna, Austria}
\affiliation{Institute for Quantum Optics and Quantum Information (IQOQI), Boltzmangasse 3 1090 Vienna, Austria}
\affiliation{Centre for Engineered Quantum Systems, School of Mathematics and Physics, The University of Queensland, St Lucia, QLD 4072, Australia}
\author{Časlav Brukner}
\affiliation{Faculty of Physics, University of Vienna, Boltzmanngasse 5 1090 Vienna, Austria}
\affiliation{Institute for Quantum Optics and Quantum Information (IQOQI), Boltzmangasse 3 1090 Vienna, Austria}
\date{\today}

\begin{abstract}
In a scenario where two parties share, act on and exchange some physical resource, the assumption that the parties' actions are ordered according to a definite causal structure yields constraints on the possible correlations that can be established.
We show that the set of correlations that are compatible with a definite causal order forms a polytope, whose facets define \emph{causal inequalities}.
We fully characterize this \emph{causal polytope} in the simplest case of bipartite correlations with binary inputs and outputs. We find two families of nonequivalent causal inequalities; both can be violated in the recently introduced framework of \emph{process matrices}, which extends the standard quantum formalism by relaxing the implicit assumption of a fixed causal structure. Our work paves the way to a more systematic investigation of causal inequalities in a theory-independent way, and of their violation within the framework of process matrices.
\end{abstract}

\maketitle

\section{Introduction}

In our common understanding of the world, we typically perceive events as being embedded in some causal structure, where events happening earlier can influence events happening later but not vice versa. Correlations can be established in such a picture by physical systems that may be shared or exchanged by different parties, and which may be used to communicate or convey causal influences.

It is well known, however, that this view is challenged by quantum correlations: Bell's theorem~\cite{bell64} shows for instance that these conflict with Reichenbach's common cause principle~\cite{reichenbach1956direction, sep-physics-Rpcc}, so that quantum mechanics forces us to generalize the notion of causal influence~\cite{laskey_quantum_2007, leifer_towards_2013, wood15, cavalcanti14, henson14, pienaar_graph-separation_2015}. Another implication of this picture is that if one assumes that the parties interact only once with the physical medium, then only one-way influences (\ie, one-way signaling) are possible, which restricts---independently of any assumptions on the physics of the involved systems---the possible correlations that can be observed.

But is this view that events should comply with a definite causal structure, and causal influences can only be unidirectional, necessary in any physical theory? Or could one envisage theories where the causal relations between events are not necessarily well defined~\cite{hardy2005probability, hardy2007towards}? To answer these questions, Oreshkov, Costa and Brukner developed the framework of \emph{process matrices} as an extension of quantum theory, where the assumption of a fixed causal structure is relaxed~\cite{oreshkov12}. Process matrices describe the physical resource that allows different parties to establish correlations. Oreshkov \emph{et al.} showed that certain so-called \emph{causally nonseparable process matrices} indeed do not comply with a definite causal structure.

The incompatibility of a certain causally nonseparable process matrix with a definite causal structure was proven in Ref.~\cite{oreshkov12} by its ability to generate correlations that are incompatible with a definite causal order, as demonstrated by the violation of a so-called \emph{causal inequality}. This can be tested in a device-independent manner, by just looking at the statistics observed in an experiment.
It was recently shown that causal nonseparability could also be detected in a device-dependent manner by using so-called \emph{causal witnesses}~\cite{araujo15}.
This approach is more powerful as it can detect all causally nonseparable process matrices, while not all causally nonseparable process matrices can violate a causal inequality~\cite{araujo15,oreshkov15}. Furthermore, physical implementations of certain (multipartite) causally nonseparable process matrices, and of corresponding causal witnesses that detect their causal nonseparability, have been proposed~\cite{chiribella09,araujo15,oreshkov15} and even realized experimentally~\cite{procopio_experimental_2014}, while it is still not known whether there actually exist any physically realizable process that violates a causal inequality.
Nevertheless, the device-independent approach is still of interest as it relaxes the requirement to trust the functioning and the operations implemented by one's devices in an experiment. It is furthermore also theory-independent: causal inequalities can in principle be tested, and correlations with no definite causal order can be identified whatever the description of the physical resource is---whether we use the process matrix framework or any other theory to be discovered in the future. A related open question is whether the ability to violate causal inequalities can---in analogy with Bell nonlocality~\cite{brunner14}---be exploited as a resource, just like causally non-separable process matrices provide advantages for information-theoretical~\cite{chiribella12} and computational~\cite{araujo14} tasks.

Our paper aims at providing a better understanding of the device-independent characterization of correlations that are compatible with a definite causal order or not. 
We show that bipartite correlations with a definite causal order form a convex polytope, whose facets correspond to causal inequalities (Section~\ref{sec:causal_correlations}). 
We characterize this \emph{causal polytope} in the simplest scenario where the two parties observe correlations with binary inputs and outputs, which gives us two families of new causal inequalities.
We then investigate their possible violation in the framework of process matrices, and find that these can indeed be violated (Section~\ref{sec:violations}). This provides an example of \emph{``noncausal''} process matrix correlations in a simpler scenario than that considered in Ref.~\cite{oreshkov12}, where one party had two input bits, or in Refs.~\cite{baumeler13,baumeler14}, where more parties were involved.

\section{Correlations \\ with definite causal order} \label{sec:causal_correlations}

\subsection{``Causal correlations''}

We consider an experiment with two parties, Alice and Bob, each of them having control over some closed laboratory.
They both open their lab, let some physical system in, interact with it and send a physical system out, only once during each run of the experiment.
Alice and Bob are given some classical inputs labeled by $x$ and $y$, and return some classical outputs $a$ and $b$, respectively. We assume that all inputs and outputs have a finite number of possible values. The correlation that Alice and Bob establish in such an experiment is described by the joint conditional probability distribution $p(a,b|x,y)$.

In a situation where at each run of the experiment Alice's events precede Bob's events (denoted $A \prec B$), Alice could send her input and output to Bob, but not vice versa; hence, there cannot be any signaling from Bob to Alice, and their correlation, which we shall denote in this case $p^{A \prec B}$, must therefore satisfy
\begin{equation}
\forall \, x,y,y',a, \quad p^{A \prec B}(a|x,y) =  p^{A \prec B}(a|x,y') \,, \label{eq:no_sig_to_A}
\end{equation}
with $p^{A \prec B}(a|x,y^{(\prime)}) = \sum_b p^{A \prec B}(a,b|x,y^{(\prime)})$.
Similarly, in a situation where Bob's events precede Alice's ($B \prec A$), their correlation $p^{B \prec A}$ must satisfy the no-signaling-to-Bob constraint
\begin{equation}
\forall \, x,x',y,b, \quad p^{B \prec A}(b|x,y) =  p^{B \prec A}(b|x',y) \,, \label{eq:no_sig_to_B}
\end{equation}
with $p^{B \prec A}(b|x^{(\prime)},y) = \sum_a p^{B \prec A}(a,b|x^{(\prime)},y)$. Note that non-signaling correlations satisfy both Eqs.~\eqref{eq:no_sig_to_A} and~\eqref{eq:no_sig_to_B}, and are compatible with both causal orders $A \prec B$ and $B \prec A$.
More generally, if the correlation is compatible with the causal order $A \prec B$ with probability $q$, and with $B \prec A$ with probability $1-q$, then the correlation will be of the form
\begin{gather}
p(a,b|x,y) = q \, p^{A \prec B}(a,b|x,y) + (1{-}q) \, p^{B \prec A}(a,b|x,y) \,. \label{def:csep}
\end{gather}

Following Refs.~\cite{brukner14b,oreshkov15,araujo15}, we call the bipartite probability distribution $p(a,b|x,y)$ (or the correlation it describes, equivalently) \emph{``causal''} if it can be written as in Eq.~\eqref{def:csep}, with $q \in [0,1]$ and $p^{A \prec B}$ and $p^{B \prec A}$ valid (\ie, nonnegative and normalized) probability distributions satisfying Eqs.~\eqref{eq:no_sig_to_A} and~\eqref{eq:no_sig_to_B}, respectively. Causal correlations are those that can be obtained in a situation where every run of the experiment is compatible with a definite causal order ($A \prec B$ or $B \prec A$), which may however vary for each run, and is only determined probabilistically. Note that the decomposition~\eqref{def:csep} is in general not unique, as non-signaling contributions can be included in either $p^{A \prec B}$ or $p^{B \prec A}$.

\subsection{Causal polytopes and causal inequalities}
     
Correlations that are compatible with the causal order $A \prec B$ satisfy nonnegativity ($p^{A \prec B}(a,b|x,y) \geq 0$ $\forall x,y,a,b$) and normalization ($\sum_{a,b} p^{A \prec B}(a,b|x,y) = 1$ $\forall x,y$) constraints, together with the no-signaling-to-Alice constraint~\eqref{eq:no_sig_to_A}. As these constitute a finite number of linear constraints on a bounded probability space\footnote{The probability space can for instance be understood geometrically as the set of points in a high enough dimensional space, whose coordinates are the values $p(a,b|x,y)$. Clearly, the nonnegativity and normalization constraints make it bounded.}, it follows that the set of correlations $p^{A \prec B}$ is a (convex) polytope~\cite{ziegler95}. 
Similarly, the set of correlations $p^{B \prec A}$ that are compatible with the causal order $B \prec A$ is also a polytope.
Now, according to Eq.~\eqref{def:csep}, the set of causal correlations is simply the convex hull of the sets of correlations $p^{A \prec B}$ and $p^{B \prec A}$, and is therefore itself a polytope, which we call the \emph{causal polytope}.

By construction, the extremal points of the causal polytope are extremal points of either the polytope of $p^{A \prec B}$ correlations, or of the polytope of $p^{B \prec A}$ correlations (or of both polytopes); in Appendix~\ref{app:characterization} we show that these correspond to deterministic correlations compatible with either causal order (or both, in the case of nonsignaling correlations). From this ``$\mathcal V$-representation'' of the causal polytope in terms of its vertices, for a given number of inputs and outputs, one can in principle determine its equivalent ``$\mathcal H$-representation'' in terms of its facets~\cite{ziegler95} (although in practice, this is a hard problem to solve when the number of inputs and outputs increase). Some of its facets are trivial, in the sense that they only correspond to the nonnegativity constraints $p(a,b|x,y) \geq 0$; its other, nontrivial facets define so-called \emph{causal inequalities}~\cite{oreshkov12}---inequalities that are satisfied by any causal correlation.

The above characterization hints of course at a strong analogy with Bell inequalities, which may be obtained as facets of the ``local polytope''~\cite{bell64,pitowsky89,brunner14} (or may not; in the same way that not all Bell inequalities are facets of the local polytope, not all causal inequalities are facets of the causal polytope, as they can also correspond to some external hyperplanes\footnote{E.g., one can check that the original causal inequality of Ref.~\cite{oreshkov12} is not a facet of the causal polytope for 1 input bit for Alice, 2 input bits for Bob, and 1 output bit for each of them (the $320$ vertices of the causal polytope that saturate that inequality, out of $5056$ vertices, only span a $21$-dimensional affine subspace, while a facet of this $24$-dimensional polytope should have dimension $23$). Likewise, the causal inequalities~(\ref{eq:proj_causal_pp}--\ref{eq:proj_causal_mm}) below are not facets of the causal polytope for binary inputs and outputs (they are only facets of its projection onto the plane considered in Subsection~\ref{subsec_boundary}).}).
Causal inequalities are written as linear combinations of the conditional probabilities $p(a,b|x,y)$, constrained by some \emph{``causal bounds''}. They can also be translated in the language of \emph{``causal games''} by considering for instance the linear combination to define the score, or possibly the probability of success (for some specific distribution of inputs), of some game. They can be tested experimentally in a device-independent way---\ie, by just considering the observed statistics, without making any assumptions on the functioning of the physical devices used in the experiment: a violation of a causal inequality guarantees that the observed correlation is incompatible with a definite causal order---or, in short, is \emph{noncausal}.

\subsection{The simplest causal polytope}
     
To illustrate the previous discussion, we now turn to the characterization of the simplest nontrivial causal polytope.
Note that causal inequalities can only be nontrivial if each party has nontrivial inputs and outputs---\ie, if they can take at least two different values. Indeed, if one party only has trivial inputs or outputs, then clearly either~\eqref{eq:no_sig_to_A} or~\eqref{eq:no_sig_to_B} holds, so that any correlation is compatible with a definite causal order.

Hence, the simplest candidate for a nontrivial causal polytope is the case with a single bit of input and output for each of the two parties\footnote{\label{footnote_fewer_outputs} Actually, one also has a nontrivial causal polytope in a scenario where one of Alice and/or Bob's input yields a binary output, while the other always gives the same output (or has no output, equivalently). In such a case the only nontrivial causal inequalities are of the LGYNI type, Eq.~\eqref{eq:LGYNI0} or~\eqref{eq:LGYNI} (note on the other hand that the corresponding local polytope is trivial~\cite{pironio05}). For simplicity however, we choose to impose throughout the paper that all inputs should have the same number of outputs.} (which we shall denote by $0$ or $1$), reminiscent of the scenario considered by Clauser--Horne--Shimony--Holt (CHSH) in the case of nonlocality~\cite{chsh69}. We generated the list of its $112$ deterministic vertices (see Appendix~\ref{app:characterization}), and enumerated its $48$ facets using the software \texttt{lrs}~\cite{lrs}.

16 of these facets are trivial, corresponding to the nonnegativity constraints $p(ab|xy) \ge 0$. 
By relabeling the inputs and outputs, the 32 remaining, non-trivial facets can be grouped in two non-equivalent families of causal inequalities: 16 facets are relabelings of the inequality
\begin{equation}
  \label{eq:gyni0}
  \frac{1}{4} \sum_{x,y,a,b} \delta_{a,y} \, \delta_{b,x}\ p(a,b|x,y) \ \le \ \frac12 \,,
\end{equation}
where $\delta_{i,j}$ is the Kronecker delta,
while the last 16 facets are relabelings of the inequality
\begin{equation}
  \label{eq:LGYNI0}
  \frac{1}{4} \sum_{x,y,a,b} \delta_{x(a \oplus y),0} \, \delta_{y(b \oplus x),0}\ p(a,b|x,y) \ \le \ \frac34 \,,
\end{equation}
where $\oplus$ denotes addition modulo 2.

The causal inequality~\eqref{eq:gyni0} can be interpreted as a bound on the maximal probability of success for a bipartite \emph{``guess your neighbor's input''} (GYNI) game~\cite{almeida10} with uniform input bits $x,y$ (such that $p(x,y) = \frac{1}{4}$), where Alice and Bob's task is to guess each other's input, \ie, to output $a=y$ and $b=x$. 
Implicitly assuming uniform input bits\footnote{Note that the assumption of uniform inputs is only necessary to justify the shorthand notation $p({a = y}, b = x)$ for the left hand side of Eq.~\eqref{eq:gyni0}, and to interpret it as the success probability for the GYNI game. Whether an inequality written as a combination of conditional probabilities (like~\eqref{eq:gyni0} or~\eqref{eq:LGYNI0} for instance) defines a causal inequality or not depends of course in no way on the distribution of inputs.}, inequality~\eqref{eq:gyni0} can indeed be written in a more compact form as
\begin{equation}
  \label{eq:gyni}
  p_\text{GYNI} \ := \ p({a = y}, b = x) \ \le \ \frac{1}{2} \,.
\end{equation}
This causal bound on the probability of success $p_\text{GYNI}$ can easily be understood: assuming that the correlation is compatible with the causal order $A \prec B$, Alice cannot know anything about Bob's input bit and can therefore only make a random guess, so that $p(a=y) = \frac12$ and therefore $p({a = y}, b = x) \le \frac{1}{2}$; a similar reasoning holds for the causal order $B \prec A$, and a convex mixture cannot increase the bound on $p_\text{GYNI}$.

Similarly, the causal inequality~\eqref{eq:LGYNI0} can be interpreted as a bound on the maximal probability of success for what we shall call a \emph{``lazy GYNI''} (LGYNI) game, still with uniformly random input bits, where Alice and Bob's task is now to guess each other's input only when their respective input is $1$ (for an input $0$, their output can be arbitrary). 
Implicitly assuming uniform input bits, inequality~\eqref{eq:LGYNI0} can then also be written in a more compact form as
\begin{equation}
  \label{eq:LGYNI}
  p_\text{LGYNI} \ := \ p\big(x(a \oplus y)=0, y(b \oplus x)=0\big) \ \le \ \frac34 \,.
\end{equation}
This causal bound on the probability of success $p_\text{LGYNI}$ can also easily be understood with a similar reasoning as above (taking into account that Alice for instance is only asked to guess Bob's input half of the time, when her input is $1$).

\section{Process matrix correlations \\ with no definite causal order} \label{sec:violations}

In this section we study the violation of our simplest causal inequalities in the framework of \emph{process matrices}, introduced recently by Oreshkov, Costa and Brukner~\cite{oreshkov12}. Let us first start with a brief overview of this framework.

\subsection{The process matrix framework}

The basic assumption of the framework is that quantum theory correctly describes what happens \emph{locally} in Alice and Bob's laboratories; however, no assumption is being made about the \textit{global} causal structure in which the parties operate.

More specifically, it is assumed that Alice and Bob can perform any operation allowed by the standard formulation of quantum theory, as described by quantum \emph{instruments}~\cite{davies70} from some input Hilbert spaces $\mathcal H^{A_I}$ and $\mathcal H^{B_I}$ (for Alice and Bob, respectively) to some output Hilbert spaces $\mathcal H^{A_O}$ and $\mathcal H^{B_O}$. An instrument is a set of completely positive (CP), trace non-increasing maps from $\mathcal L(\mathcal H^{X_I})$ to $\mathcal L(\mathcal H^{X_O})$ (for $X = A, B$), where $\mathcal L(\mathcal H^{X_I})$ and $\mathcal L(\mathcal H^{X_O})$ are the spaces of linear operators over the Hilbert spaces $\mathcal H^{X_I}$ and $\mathcal H^{X_O}$. Each CP map of a given instrument is associated with a given classical output, which we shall again denote by $a$ and $b$ for Alice and Bob, and all CP maps of an instrument must sum up to a trace-preserving map. The various instruments that the parties can choose to apply shall be labeled by some classical ``inputs'' $x$ and $y$.

Using the Choi-Jamio\l{}kowski (CJ) isomorphism~\cite{jamio72,choi_completely_1975}, one can represent Alice's maps as some operators\footnote{Throughout the paper, superscripts on operators refer to the Hilbert space they act on.} $M_{a|x}^{A_IA_O}$ on the tensor product space $\mathcal{L}(\mathcal{H}^{A_I} \otimes \mathcal{H}^{A_O})$.
The conditions for the collection of operators $\{M_{a|x}^{A_IA_O}\}_a$ (for some fixed input $x$) to be a valid instrument translate to
\begin{eqnarray}
M_{a|x}^{A_IA_O} \geq 0 \quad \forall \, a \mathand \tr_{A_O} \sum_a M_{a|x}^{A_IA_O} = \id^{A_I} , \quad \label{eq:valid_instrument}
\end{eqnarray}
where $\tr_{A_O}$ denotes the partial trace over Alice's output system, and $\id^{A_I}$ is the identity operator in Alice's input Hilbert space. Similarly, Bob's maps can be represented as operators $M_{b|y}^{B_IB_O}$ on $\mathcal{L}(\mathcal{H}^{B_I} \otimes \mathcal{H}^{B_O})$, and a collection of operators $\{M_{b|y}^{B_IB_O}\}_b$ (for some fixed input $y$) must satisfy analogous constraints to Eq.~\eqref{eq:valid_instrument} to be a valid instrument.

As shown in Ref.~\cite{oreshkov12}, the assumption of local consistency with quantum theory implies that the probability $p(a,b|x,y)$ of observing the classical outputs $a,b$ for a choice of instruments labeled by $x,y$ is a bilinear function of Alice and Bob's maps, which can be written as
\begin{equation}
  \label{eq:probw}
  p(a,b|x,y) = \tr\big[ (M_{a|x}^{A_IA_O} \otimes M_{b|y}^{B_IB_O}) \cdot W\big] \, ,
\end{equation}
for some hermitian matrix $W \in \mathcal{L}(\mathcal{H}^{A_I} \otimes \mathcal{H}^{A_O} \otimes \mathcal{H}^{B_I} \otimes \mathcal{H}^{B_O})$.
Requiring that the probabilities given by~\eqref{eq:probw} are nonnegative and normalized for all possible choice of quantum operations (including operations involving possibly entangled ancillary systems) imposes some restrictions on the possible $W$ matrices~\cite{oreshkov12}. As shown in Ref.~\cite{araujo15}, these constraints can be expressed as follows:
\begin{subequations}\label{eq:valid_W}
\begin{gather}
 W \ge 0 \, , \\[1mm]
 \tr W = d_{A_O} \, d_{B_O} \, , \\[1mm]
 {}_{B_IB_O}W = {}_{A_OB_IB_O}W \, , \\[1mm]
 {}_{A_IA_O}W = {}_{A_IA_OB_O}W \, , \\[1mm]
 W = {}_{B_O}W + {}_{A_O}W - {}_{A_OB_O}W \, ,
\end{gather}
\end{subequations}
where the last three conditions are written using the operation $_X\cdot$ defined by
\begin{equation}
 _X W = \frac{\id^{X}}{d_X} \otimes \tr_X W
\end{equation}
for $X = A_I, A_O, B_I, B_O$, with $\id^{X}$ and $\tr_X$ denoting the identity operator and the partial trace over the Hilbert space $\mathcal H^X$, respectively, and $d_X$ denoting its dimension.

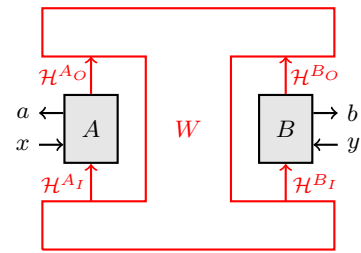
\begin{figure}[htpc]
  \centering
  \begin{tikzpicture}[scale=1.6]
		\node[draw, thick, rectangle,minimum width=0.7cm,minimum height=0.9cm, fill=white!80!gray] (S) at (-0.8,0) {$A$};
		\node[red,thick, minimum width=0.7cm,minimum height=0.9cm] (W) at (0,0) {$W$};
		\node[draw, thick, rectangle,minimum width=0.7cm,minimum height=0.9cm, fill=white!80!gray] (T) at (0.8,0) {$B$};
                \draw[red, thick] (-1.2,-1) -- (1.2,-1) -- (1.2,-0.6) -- (0.35,-0.6) -- (0.35,0.6) -- (1.2,0.6) -- (1.2,1) -- (-1.2,1) -- (-1.2, 0.6) -- (-0.35,0.6) -- (-0.35,-0.6) -- (-1.2,-0.6) -- (-1.2,-1);
                \draw[->, red, thick] (S) -- (-0.8,0.6);
                \draw[<-, red, thick] (S) -- (-0.8,-0.6);
                \draw[->, red, thick] (T) -- (0.8,0.6);
                \draw[<-, red, thick] (T) -- (0.8,-0.6);
		\node[red] (SO) at ([shift={(-0.22cm,0.45cm)}]S) {\footnotesize $\Hi^{A_O}$};
		\node[red] (SI) at ([shift={(-0.22cm,-0.45cm)}]S) {\footnotesize $\Hi^{A_I}$};
		\node[red] (TO) at ([shift={(0.25cm,0.45cm)}]T) {\footnotesize $\Hi^{B_O}$};
		\node[red] (TI) at ([shift={(0.25cm,-0.45cm)}]T) {\footnotesize $\Hi^{B_I}$};
		\draw[->, thick] (S.150) -- ++(-0.2,0) node [left] {$a$};
		\draw[<-, thick] (S.210) -- ++(-0.2,0) node [left] {$x$};
		\draw[->, thick] (T.30) -- ++(0.2,0) node [right] {$b$};
		\draw[<-, thick] (T.-30) -- ++(0.2,0) node [right] {$y$};
  \end{tikzpicture}
  \caption{A process matrix $W$ represents the physical resource which connects Alice's ($\Hi^{A_O}$) and Bob's ($\Hi^{B_O}$) output Hilbert spaces to their input Hilbert spaces ($\Hi^{A_I}$, $\Hi^{B_I}$) in such a way that what happens in Alice and Bob's labs is locally consistent with quantum theory~\cite{oreshkov12}. Process matrices generalize in particular quantum states and quantum channels.}
  \label{fig:w}
\end{figure}

Operators $W$ that satisfy these conditions are called \textit{process matrices}. They represent the most general way to ``connect'' the output spaces $\Hi^{A_O} \otimes \Hi^{B_O}$ to the input spaces $\Hi^{A_I} \otimes \Hi^{B_I}$ (see Fig.~\ref{fig:w}) in a way that is locally consistent with quantum theory. While these conditions do not impose a global causal order \emph{a priori} and therefore allow in general for two-way signaling, the nonnegativity and normalization conditions on the probabilities guarantee that no logical paradoxes, like the grandfather paradox  for instance~\cite{schachner33,barjavel43}, appear.
In the following we will refer to correlations of the form~\eqref{eq:probw}, with Alice and Bob's instruments satisfying Eq.~\eqref{eq:valid_instrument} (together with its analogous form for Bob) and $W$ satisfying the conditions~\eqref{eq:valid_W}, as \textit{process matrix correlations}.

\medskip

\subsection{Violation of the simplest causal inequalities \newline by process matrix correlations}

It was shown in Ref.~\cite{oreshkov12} that certain process matrices\footnote{A necessary condition for a process matrix to allow for a causal inequality violation is that it is \emph{causally nonseparable}~\cite{oreshkov12}---\ie, that it is itself incompatible with a definite causal order. In the multipartite case this is known however not to be a sufficient condition~\cite{araujo15,oreshkov15}. It remains an open question whether there can be bipartite causally nonseparable process matrices that only generate causal correlations.} could generate correlations with no definite causal order. A specific process matrix and specific instruments were indeed found, which violate a particular causal inequality with one input bit for Alice and two for Bob, and one output bit for each.
Remarkably, one of Bob's input bits could be used to distinguish some runs of the experiment where signaling happened in one direction, and some runs where it happened in the other direction.
It remained an open question, whether this special input bit for Bob was necessary to obtain noncausal correlations in the process matrix framework, or whether any simpler causal inequality (with fewer inputs) could be violated.
Here we answer this question positively, by exhibiting violations of both our GYNI and LGYNI inequalities~(\ref{eq:gyni}, \ref{eq:LGYNI}) by process matrix correlations.

\medskip

Let us start with a simple example with two-dimensional input and output systems---``qubits''---for Alice and Bob (\ie, $d_{A_I} = d_{A_O} = d_{B_I} = d_{B_O} = 2$). One can check that the matrix
\begin{equation}
  \label{eq:wsimple}
  W = \frac{1}{4} \left[ \id^{\otimes 4} + \frac{Z^{A_I} Z^{A_O} Z^{B_I} \id^{B_O} + Z^{A_I} \id^{A_O} X^{B_I} X^{B_O}}{\sqrt{2}} \right] \, , 
\end{equation}
where $Z$ and $X$ are the Pauli matrices and where tensor products are implicit, satisfies the constraints~\eqref{eq:valid_W}, so that it defines a valid process matrix. We choose Alice and Bob's operations to be the same, defined by
\begin{align}
 M_{0|0}^{A_IA_O} &= M_{0|0}^{B_IB_O} = 0 \, , \label{def:M00} \\
 M_{1|0}^{A_IA_O} &= M_{1|0}^{B_IB_O} = 2 \, \proj{\Phi^+} \, , \label{def:M10} \\
 M_{0|1}^{A_IA_O} &= M_{0|1}^{B_IB_O} = \proj{0} \otimes \proj{0} \, , \label{def:M01} \\
 M_{1|1}^{A_IA_O} &= M_{1|1}^{B_IB_O} = \proj{1} \otimes \proj{0} \, , \label{def:M11}
\end{align}
with $\{\ket{0}, \ket{1}\}$ denoting the computational basis (\ie, the eigenbasis of $Z$), and $\ket{\Phi^{+}} : = (\ket{00}+\ket{11})/\sqrt{2}$. These indeed satisfy~\eqref{eq:valid_instrument}, and thus constitute valid instruments. These operations can be interpreted as follows: when their input is $0$, Alice and Bob simply transmit their incoming physical system, untouched ($2 \, \proj{\Phi^+}$ being indeed the CJ representation of an identity channel), and output the value $1$; when their input is $1$, Alice and Bob perform a measurement in the $Z$ basis, whose result defines their classical output, and send out the fixed state $\proj{0}$.
With these definitions, one can calculate the success probabilities of the GYNI and LGYNI games using Eqs.~(\ref{eq:gyni}, \ref{eq:LGYNI})---or more explicitly~(\ref{eq:gyni0}, \ref{eq:LGYNI0})---and Eq.~\eqref{eq:probw}. One finds
    \begin{align}
p_{\text{GYNI}} &\, = \, \frac5{16}\Big(1+\frac1{\sqrt{2}}\Big) \, \approx \, 0.5335 \, > \, \frac12 \,, \\
p_{\text{LGYNI}} &\, = \, \frac5{16}\Big(1+\frac1{\sqrt{2}}\Big) + \frac14 \, \approx \, 0.7835 \, > \, \frac34 \,,
    \end{align}
which indeed violate the causal inequalities~(\ref{eq:gyni}, \ref{eq:LGYNI}).

\medskip

One may now wonder, what the largest possible violation of these two causal inequalities by process matrix correlations is. To optimize the violations for some input and output Hilbert spaces of a given dimension, we used a See-Saw algorithm inspired by that of Werner and Wolf~\cite{wernerwolf}, as described in Appendix~\ref{app:seesaw}.
Note that because the optimization problem is nonconvex, the algorithm is not guaranteed to converge to a global maximum. Nevertheless, for small enough dimensions (at least, for qubits), the repeatability of our results for different random starting points of the algorithm makes us confident that we indeed found the global maxima.
From our numerical results, we thus conjecture that the maximal violations of our causal inequalities achievable with qubit systems are
    \begin{align}
p_{\text{GYNI}}^{\text{max}, d=2} & \, \approx \, 0.5694 \, > \, \frac12 \,, \\
p_{\text{LGYNI}}^{\text{max}, d=2} & \, \approx \, 0.8194 \,=\, p_{\text{GYNI}}^{\text{max}, d=2} + \frac14 \, > \, \frac34 \,.
    \end{align}
In Appendix~\ref{app:maximal_qubit} we give an analytical description of the process matrices that reach these values.

Going to larger dimensions, we found that the maximal value of $p_{\text{GYNI}}$ could increase, as summarized in Table~\ref{tab:violations} for dimensions up to $5$; however, we did not find any larger value for $p_{\text{LGYNI}}$ than $p_{\text{LGYNI}}^{\text{max}, d=2}$ above, which---provided our See-Saw algorithm did find the global maxima---reveals some fundamental difference between the two inequalities, despite their similarities (and in addition to the fact that contrary to GYNI, the outputs corresponding to certain inputs are irrelevant in the LGYNI game; see also footnote~\ref{footnote_fewer_outputs}).
It remains an open question, which values are the true ``Tsirelson bounds''~\cite{tsirelson80} for these two causal inequalities, in the sense of the largest possible values for $p_\text{GYNI}$ and $p_\text{LGYNI}$ reachable with quantum process matrices of any dimension.

\begin{table}[htpc]
\vspace{5mm}
\begin{tabular}{c|cccc}
$d$ & $2$ & $3$ & $4$ & $5$ \\ \hline
\ $p_{\text{GYNI}}^{\text{max}, d}$ \ & \ 0.5694 \ & \ 0.6104 \ & \ 0.6201 \ & \ 0.6218
\end{tabular} 
\caption{Maximal values of $p_{\text{GYNI}}$ found through numerical optimization, as a function of the dimension of Alice and Bob's input and output Hilbert spaces $d = d_{A_{I}} = d_{A_{O}} =d_{B_{I}} = d_{B_{O}}$.}
\label{tab:violations}
\end{table}

\subsection{Boundary of the set \\ of process matrix correlations} \label{subsec_boundary}

\begin{figure}[tpc]
\begin{tikzpicture}
\begin{axis}[ymin=-0.1,ymax=1.1,xmin=0,xmax=1,area style,axis equal,xlabel={$p(a\!=\!y)$},ylabel={$p(b\!=\!x)$}]
\addplot[draw=black,dashed] coordinates {(0,0) (0,1) (1,1) (1,0) (0,0)};
\addplot[draw=purple,fill=purple!30!white] file {boundary4pp};
\addplot[draw=purple,fill=purple!30!white] file {boundary4pm};
\addplot[draw=purple,fill=purple!30!white] file {boundary4mp};
\addplot[draw=purple,fill=purple!30!white] file {boundary4mm};
\addplot[draw=green,fill=green!30!white] file {boundary3pp};
\addplot[draw=green,fill=green!30!white] file {boundary3pm};
\addplot[draw=green,fill=green!30!white] file {boundary3mp};
\addplot[draw=green,fill=green!30!white] file {boundary3mm};
\addplot[draw=red,fill=red!30!white] file {boundary2pp};
\addplot[draw=red,fill=red!30!white] file {boundary2pm};
\addplot[draw=red,fill=red!30!white] file {boundary2mp};
\addplot[draw=red,fill=red!30!white] file {boundary2mm};
\addplot[draw=blue,thick,fill=blue!30!white] coordinates {(0,1/2) (1/2,1) (1,1/2) (1/2,0) (0,1/2)};
\end{axis}
\end{tikzpicture}
\caption{Projection of the probability space for binary inputs and outputs onto the plane $\big( p(a\!=\!y), p(b\!=\!x) \big)$ (see main text). The causal polytope is projected onto the blue diamond, delimited by the causal inequalities~(\ref{eq:proj_causal_pp}--\ref{eq:proj_causal_mm}). The red, green and purple regions correspond to process matrix correlations that are reachable with input and output Hilbert spaces of dimensions $d = 2, 3$ and $4$, respectively. The outer dashed square delimits all valid probabilities $p(a\!=\!y), p(b\!=\!x) \in [0,1]$. Its upper right corner corresponds for instance to a correlation such that Alice's output is always equal to Bob's input ($a\!=\!y$) and Bob's output is always equal to Alice's input ($b\!=\!x$), which requires perfect 2-way signaling and violates Eqs.~\eqref{eq:gyni}, \eqref{eq:LGYNI} and~\eqref{eq:proj_causal_pp} up to their algebraic maximum. This correlation may somehow be thought of as being analogous to the Popescu-Rohrlich (PR) box considered in the context of nonlocal correlations~\cite{popescu94} (one difference, however, is that this correlation is deterministic, while the PR box correlations are not).}
\label{fig:boundary}
\end{figure}
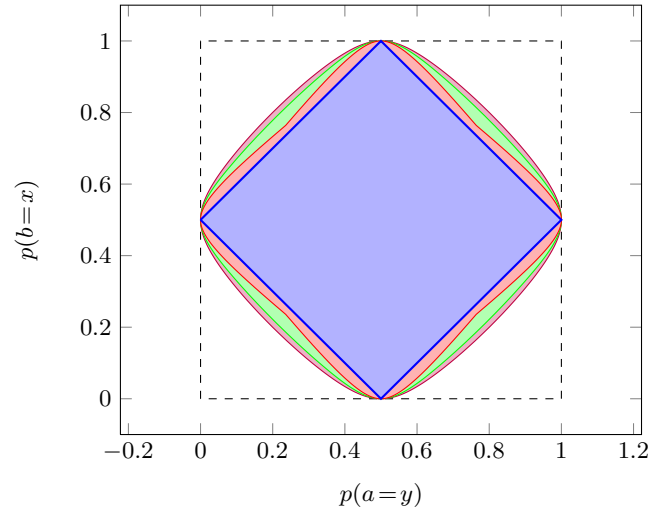

To finish with, let us picture the set of process matrix correlations vs that of causal correlations. In order to visualize the two, we shall project them onto the plane with coordinates $\big( p(a\!=\!y), p(b\!=\!x)\big)$, where we again implicitly assume uniformly random inputs for ease of notations\footnote{Without assuming uniformly random inputs, $p(a\!=\!y)$ and $p(b\!=\!x)$ in Eqs.~(\ref{eq:proj_causal_pp}--\ref{eq:proj_causal_mm}) and in Figure~\ref{fig:boundary} should be replaced by $\frac 14 \sum_{x,y,a,b} \delta_{a,y} \, p(a,b|x,y)$ and $\frac 14 \sum_{x,y,a,b} \delta_{b,x} \, p(a,b|x,y)$, respectively.}. In this plane the complementarity between the two directions of signaling, from Alice to Bob and from Bob to Alice, is conspicuous; the projected causal polytope is bounded here by the four causal inequalities\footnote{Note that inequality~\eqref{eq:proj_causal_pp} looks quite similar to Oreshkov \emph{et al.}'s original causal inequality~\cite{oreshkov12} (which motivated us in particular to write it with the $\frac 12$ factors). Oreshkov \emph{et al.}'s inequality includes some additional conditioning on a second input bit for Bob. Averaging that inequality with its equivalent version where Bob's second input bit is flipped yields inequality~\eqref{eq:proj_causal_pp}. Similarly, Brukner's ``Tsirelson-like bound'' for Oreshkov \emph{et al.}'s inequality~\cite{brukner14} (for a limited set of possible instruments) also yields the same Tsirelson-like bound $\frac{1+1/\sqrt{2}}{2} \simeq 0.8536$ for inequality~\eqref{eq:proj_causal_pp} (with the same restriction); there remains a large gap between that bound and the lower bounds we obtained numerically for dimensions $2, 3$ and $4$.} (see Figure~\ref{fig:boundary})
  \begin{align}
   \frac 12 \, p(a=y) + \frac 12 \, p(b=x) & \, \le \, \frac 34 \, , \label{eq:proj_causal_pp} \\[1mm]
   \frac 12 \, p(a=y) - \frac 12 \, p(b=x) & \, \le \, \frac 14 \, , \label{eq:proj_causal_pm} \\[1mm]
   -\frac 12 \, p(a=y) + \frac 12 \, p(b=x) & \, \le \, \frac 14 \, , \label{eq:proj_causal_mp} \\[1mm]
   -\frac 12 \, p(a=y) - \frac 12 \, p(b=x) & \, \le \, -\frac 14 \, . \label{eq:proj_causal_mm}
  \end{align}

  To obtain a lower bound for the boundary of the set of process matrix correlations, we again used the See-Saw algorithm described in Appendix~\ref{app:seesaw} to maximize quantities of the form $\alpha \, p(a\!=\!y) + \beta \, p(b\!=\!x)$, with various weights $\alpha, \beta$. Different runs of the algorithm gave us different lower bounds (recall that the See-Saw algorithm is not guaranteed to always find the global optimum), which we combined to obtain the bounds represented on Figure~\ref{fig:boundary} for dimensions $d = 2, 3$ and $4$ of the input and output Hilbert spaces---and which we believe are close to the actual boundaries of the process matrix correlations for these dimensions. The largest violations of the causal inequality~\eqref{eq:proj_causal_pp} we found are $\frac 12 \, p(a\!=\!y) + \frac 12 \, p(b\!=\!x) = 0.7715$ for $d=2$; $0.7892$ for $d=3$; and $0.8001$ for $d=4$.
  
  A surprising feature of the set of process matrix correlations for dimension $2$ is that it does not seem to be convex (see Figure~\ref{fig:boundary}, red region). We believe this is a true characteristic of it, not only a numerical artifact due to some failure to find global optima. Note, nevertheless, that the boundary of the set of process matrix correlations for arbitrary dimensions \textit{is} convex, as proven in Appendix~\ref{app:convex}.
  
\section{Conclusion}

We have shown that the set of correlations compatible with a definite causal order (``causal correlations'') forms a convex polytope, which we fully characterized in the simplest nontrivial bipartite scenario with binary inputs and outputs. Two nonequivalent families of causal inequalities were obtained, Eqs.~(\ref{eq:gyni}--\ref{eq:LGYNI}), for which we gave intuitive interpretations in terms of ``causal games''. These allow for a device-independent characterization of correlations with or without definite causal order, and can be tested independently of the physical theory under consideration. We exhibited in particular violations of these inequalities by process matrix correlations, which generalize standard quantum correlations. Because of their simplicity (and despite the violations we found being somewhat less intuitive), we expect these new inequalities---in particular the first one, interpreted as a ``guess your neighbor's input'' game---to become archetypical examples of causal inequalities, just like the CHSH inequality is the archetype of Bell inequalities~\cite{chsh69,brunner14}.

Our approach can be used to characterize (non)causal correlations in more complex scenarios as well. It should be noted that because of the large dimension of the probability space and the large number of vertices of the causal polytope (see Appendix~\ref{app:characterization}), the full facet enumeration problem rapidly becomes intractable in practice as the number of inputs and outputs increases beyond the simplest binary case. Nevertheless, one could adapt some of the tricks developed for the derivation of Bell inequalities (see Ref.~\cite{brunner14} for a review) to construct new causal inequalities for various scenarios of interest.
Violations of these inequalities in the process matrix framework can then be investigated using our See-Saw algorithm. An interesting question is whether it would also be possible to derive nontrivial bounds on such violations from certain information-theoretic principles~\cite{ibnouhsein2014information}, along analogous lines to the research program that aims at restricting quantum nonlocal correlations from various principles~\cite{brunner14}.

In the present paper we focused for simplicity on the bipartite case. Our work can naturally also be extended to the case of more parties. With a proper generalization of the concept of noncausal correlations (see for instance Ref.~\cite{oreshkov15}), it can also be shown that multipartite noncausal correlations form a convex polytope.
Similar techniques can be used to characterize this polytope, construct causal inequalities and investigate their possible violation. Note that a remarkable new feature in the multipartite case is that violations are also possible with ``classical process matrices''~\cite{baumeler14}.

One of the main open questions along the line of research presented here is whether it would actually be possible, in practice, to observe correlations with no definite causal order and a violation of a causal inequality. As an extension of standard quantum theory, the framework of process matrices---which does indeed predict such violations theoretically---appears as a good candidate to provide such a possibility. However, to the best of our knowledge, no practical implementation has been identified for any of the process matrices that are known to violate a causal inequality~\cite{oreshkov12,baumeler13,baumeler14} (including the ones presented here)---while in contrast, a \textit{causally nonseparable} quantum process has been recently demonstrated experimentally~\cite{procopio_experimental_2014}.
It is likely that more complex scenarios need to be considered, and a systematic investigation of causal inequalities and their violation by process matrix correlations might prove useful to find practical violations---or, should it be the case, to clarify why such violations cannot be observed in practice. Our work makes the first crucial step in this direction.

\medskip

\paragraph*{Note added.---} While finishing writing up this manuscript, we became aware that the concept of causal polytopes introduced here was also referred to (with proper reference to our work) in Ref.~\cite{oreshkov15}, where the emphasis was put on multipartite scenarios, and in Ref.~\cite{baumeler15}, where the authors also introduced, for the multipartite case as well, larger polytopes of logically consistent but possibly noncausal classical processes.

\subsection*{Acknowledgements}
We thank Ognyan Oreshkov and Michal Sedl{\'a}k for useful discussions. We acknowledge support from the French National Research Agency through the `Retour Post-Doctorants' program (ANR-13-PDOC-0026); the European Commission through a Marie Curie International Incoming Fellowship (PIIF-GA-2013-623456); the European Commission project RAQUEL (No.~323970); the Austrian Science Fund (FWF) through the Special Research Program Foundations and Applications of Quantum  Science (FoQuS), the doctoral programme CoQuS, and Individual Project (No.~2462); FQXi; the John Templeton Foundation; and the Templeton World Charity Foundation (grant TWCF 0064/AB38).

\appendix

\section{Characterization of the causal polytope}\label{app:characterization}
 
\subsection{Vertices of the causal polytope}

As explained in the main text, the causal polytope, defined as the set of causal correlations of the form~\eqref{def:csep}, is the convex hull of the polytope of correlations compatible with the causal order $A \prec B$ (which cannot signal to Alice) and the polytope of correlations compatible with the order $B \prec A$ (which cannot signal to Bob), so that its vertices are vertices of at least one of these two polytopes.

\medskip

Let us characterize the vertices of the polytope  of correlations $p^{A \prec B}$. Using Bayes' rule, we can write
  \[ p^{A \prec B}(a,b|x,y) = p(a|x,y) \, p(b|x,y,a) = p(a|x) \, p(b|x,y,a), \]
  where the last equality follows from the no-signaling-to-Alice constraint~\eqref{eq:no_sig_to_A}.
No further constraint is imposed on $p(a|x)$ and $p(b|x,y,a)$ (except that they must be valid probability distributions); these can be written as convex combinations of deterministic distributions, in the form
    \begin{align*}
  p(a|x) &\, = \, \sum_{\alpha} \, q_{\alpha} \, \delta_{a,\,\alpha(x)} \, ,  \\
  p(b|x,y,a) &\, = \, \sum_{\beta} \, q_{\beta} \, \delta_{b,\,\beta(x,y,a)} \, ,
  \end{align*}
  with $q_{\alpha}, q_{\beta} \ge 0$, $\sum_{\alpha} q_{\alpha} = \sum_{\beta} q_{\beta} = 1$, where the $\alpha$'s denote deterministic functions of Alice's input $x$, and the $\beta$'s denote deterministic functions of both Alice and Bob's inputs $x,y$ and of Alice's output $a$.
  Combining them, we get
  \begin{align*}
 p^{A \prec B}(a,b|x,y) \ &= \ \sum_{\alpha,\beta} \, q_{\alpha} \, q_{\beta} \ \delta_{a,\,\alpha(x)} \ \delta_{b,\,\beta(x,y,a)} \\
  &= \ \sum_{\alpha,\beta} \, q_{\alpha} \, q_{\beta} \ \delta_{a,\,\alpha(x)} \ \delta_{b,\,\beta_\alpha(x,y)}  \\        
  &= \ \sum_{\alpha,\beta'} \, q_{\alpha,\beta'} \ \delta_{a,\,\alpha(x)} \ \delta_{b,\,\beta'(x,y)} \,,  
\end{align*}
where $\beta_\alpha(x,y) = \beta(x,y,\alpha(x))$ and $q_{\alpha,\beta'} = q_{\alpha} \, q_{\beta'|\alpha}$ with $q_{\beta'|\alpha} = \sum_{\beta} \, \delta_{\beta_\alpha,\beta'} \, q_{\beta}$, such that $q_{\alpha,\beta'} \ge 0$ and $\sum_{\alpha,\beta'} q_{\alpha,\beta'} = 1$.
Hence, any correlation $p^{A \prec B}$ can be written as a convex combination of deterministic correlations compatible with the order $A \prec B$---which thus correspond to the vertices of the corresponding polytope of correlations $p^{A \prec B}$.

\medskip

If Alice and Bob's inputs can take $m_A$ and $m_B$ different values, respectively, and their outputs can take $k_A$ and $k_B$ values, resp., then there are $k_A^{m_A}$ different deterministic functions $\alpha(x)$ and $k_B^{m_A m_B}$ functions $\beta'(x,y)$, so that the polytope of correlations $p^{A \prec B}$ has $k_A^{m_A} \times k_B^{m_A m_B}$ vertices. Note that $k_B^{m_B}$ of the functions $\beta'(x,y)$ do not depend on $x$; hence, out of all the vertices, $k_A^{m_A} k_B^{m_B}$ are non-signaling, while the other $k_A^{m_A} (k_B^{m_A m_B} - k_B^{m_B})$ are signaling to Bob.

Similarly, the vertices of the polytope of correlations $p^{B \prec A}$ are the $k_A^{m_A m_B} \, k_B^{m_B}$ deterministic correlations compatible with the order $B \prec A$, of which $k_A^{m_A} k_B^{m_B}$ are non-signaling and are thus common to the previous polytope of correlations $p^{A \prec B}$, while the other $k_B^{m_B} (k_A^{m_A m_B} - k_A^{m_A})$ are signaling to Alice.
The vertices of the causal polytope are all the deterministic correlations\footnote{Note that the fact that the causal polytope has deterministic vertices was not trivial \emph{a priori}; this contrasts for instance with the no-signaling polytope, which has non-deterministic vertices like the PR-box~\cite{popescu94}.} compatible with either the order $A \prec B$, or the order $B \prec A$, or both---which makes a total of $k_A^{m_A} \, k_B^{m_A m_B} + k_A^{m_A m_B} \, k_B^{m_B} - k_A^{m_A} k_B^{m_B}$ vertices.

\subsection{Dimensions}

Because of the $m_A m_B$ normalization constraints $\sum_{a,b} p(a,b|x,y) = 1$, the probability space of correlations $p(a,b|x,y)$ is of dimension $m_A m_B ( k_A k_B {-} 1)$.
With the no-signaling-to-Alice and no-signaling-to-Bob constraints~(\ref{eq:no_sig_to_A},$\,\,$\ref{eq:no_sig_to_B}), the dimensions of the polytopes of correlations $p^{A \prec B}$ and $p^{B \prec A}$ are reduced to $m_A m_B ( k_A k_B {-} 1) - m_A(m_B{-}1)(k_A{-}1)$ and $m_A m_B ( k_A k_B {-} 1) - (m_A{-}1)m_B(k_B{-}1)$, respectively. However, the causal polytope---\ie, their convex hull---remains of the same dimension as the full probability space.

\subsection{For binary inputs and outputs}

In the case where both Alice and Bob's inputs and outputs take binary values, the $10$-dimensional polytopes of correlations $p^{A \prec B}$ and $p^{B \prec A}$ both have $64$ vertices, among which $16$ are non-signaling vertices common to both polytopes. The 12-dimensional causal polytope thus has $64 + 64 - 16 = 112$ different vertices.
  
We enumerated the facets of the causal polytope for binary inputs and outputs by solving the convex hull problem using the software \texttt{lrs}~\cite{lrs}. As described in the main text, we obtained 48 facets, which can be grouped into 3 families of equivalent facets (up to relabelings of inputs and outputs). Explicitly, these are
\begin{itemize}
\item 16 trivial facets of the form $p(a,b|x,y) \ge 0$ for all $x,y,a,b = 0,1$;
\item 16 facets of the GYNI type, which can be written (in the same form as~\eqref{eq:gyni}, implicitly assuming uniform input bits) as
  \begin{equation*}
  p(a \oplus \alpha_1 x \oplus \alpha_0 = y, b \oplus \beta_1 y \oplus \beta_0 = x) \ \le \ \frac{1}{2},
\end{equation*}
for all $\alpha_0, \alpha_1, \beta_0, \beta_1 = 0,1$;
\item 16 facets of the LGYNI type, which can be written (in the same form as~\eqref{eq:LGYNI}, implicitly assuming uniform input bits) as
  \begin{equation*}
  p \big( (x \oplus \alpha_1)(a \oplus \alpha_0 \oplus y)=0, (y \oplus \beta_1)(b \oplus \beta_0 \oplus x) = 0 \big) \, \le \, \frac{3}{4},
\end{equation*}
for all $\alpha_0, \alpha_1, \beta_0, \beta_1 = 0,1$.
\end{itemize}

\medskip

Note that this causal polytope for binary inputs and outputs coincides with the polytope of correlations obtained from a local model augmented with one bit of communication, as described in Ref.~\cite{bacon03}. This is because the use of just one bit of (one-way) communication is of course compatible with a definite causal order, either $A \prec B$ or $B \prec A$, and for binary inputs one bit is enough for one party to send all the information about her input to the other party. In general however, the polytopes described in Ref.~\cite{bacon03} are different from causal polytopes.

\section{See-Saw algorithm}\label{app:seesaw}

Maximizing the violation of a causal inequality over the process matrix and the instruments is a nonlinear problem, which makes it intractable directly. 
To address this problem, we used an approach inspired by the See-Saw algorithm of Werner and Wolf~\cite{wernerwolf}. The idea is that if Alice and Bob's instruments are fixed, then the combination of probabilities that enters the causal inequality is a linear function of the $W$ matrix, and maximizing it is a semidefinite programming (SDP) problem~\cite{nesterov87} that can be solved efficiently. In the same spirit, if the $W$ matrix and the instruments of one party are fixed, then the value of interest is a linear function of the instruments of the other party, and again its optimization is a SDP problem. Hence, one can try to approach the maximum violation of a causal inequality by optimizing over the process matrix and the parties' instruments in an iterative manner.
  
  More specifically, let $\omega(W,\mathcal A,\mathcal B)$ be the value taken by the combination of probabilities in the causal inequality, considered as a function of the process matrix $W$ and the sets of instruments $\mathcal A=\{M_{a|x}^{A_IA_O}\}_a$ and $\mathcal B=\{M_{b|y}^{B_IB_O}\}_b$ (in their CJ representation). We start the algorithm by generating random sets of instruments $\mathcal A_0$ and $\mathcal B_0$, and for these fixed instruments we maximize $\omega$ considered as a function of $W$, via the following SDP problem:
  \begin{gather*}
  \text{maximize}\quad\omega(W,\mathcal A_0,\mathcal B_0) \\
  \text{subject to} \\ W \geq 0 \,, \quad
  \tr W = d_{A_O}d_{B_O} \,, \\
{}_{B_IB_O}W = {}_{A_OB_IB_O}W\,, \quad
{}_{A_IA_O}W = {}_{A_IA_OB_O}W\,, \\
W = {}_{B_O}W + {}_{A_O}W - {}_{A_OB_O}W\,.
\end{gather*}
  With the optimal process matrix $W_0$ thus obtained and the fixed set of instruments $\mathcal B_0$ for Bob, we proceed to optimize $\omega$ as a function of Alice's instruments, via the SDP problem
\begin{gather*}
  \text{maximize}\quad\omega(W_0,\mathcal A,\mathcal B_0) \\
  \text{subject to} \\
  \forall \, x,a, \quad M_{a|x}^{A_IA_O} \geq 0\,, \quad
  \tr_{A_O}\sum_a M_{a|x}^{A_IA_O} = \id^{A_I}\,.
\end{gather*}
    With the optimal set of instruments $\mathcal A_0$ obtained now and the previously obtained process matrix $W_0$, we do the analogous optimization over Bob's set of instruments $\mathcal B$, and iterate the three optimization steps of the algorithm until it converges. 
    
    One can see that at each step of the algorithm the value of $\omega$ can only increase, so it is guaranteed to converge to a local maximum. One does not, however, always get the global maximum, and in practice one must repeat the algorithm several times to get a good lower bound on the maximal value of $\omega$.
    
Note that this See-Saw algorithm can of course straightforwardly be adapted to more than two parties.

\section{Maximal violations for qubits}\label{app:maximal_qubit}

The best violations of our GYNI and LGYNI causal inequalities that we found, for local dimensions $d=d_{A_I}=d_{A_O}=d_{B_I}=d_{B_O}=2$ (\ie, for ``qubits''), is reached by any convex combination
\begin{equation*}
q \, W_{\text{max}, d=2} + (1{-}q) \, W_{\text{max}, d=2}^{\prime}
\end{equation*}
(with $0 \leq q \leq 1$) of the two process matrices $W_{\text{max}, d=2}$ and $W_{\text{max}, d=2}'$ defined as
\begin{eqnarray*}
W_{\text{max}, d=2}^{(\prime)} &=& \frac14 \Big[ \id^{\otimes 4} +  a_0^{(\prime)} \, Z\id Z\id  - a_1^{(\prime)}\big(Z\id\id\id {+} \id\id Z\id\big) \\
&& \hspace{-1mm} - a_2^{(\prime)}\big(Z\id\id Z {+} \id Z Z\id\big) + a_3^{(\prime)}\big(Z\id ZZ {+} ZZZ\id\big) \\
&& \hspace{4mm} + a_4^{(\prime)}\big(Z\id XX {-} Z\id YY  {+} XXZ\id {-} YYZ\id\big)  \Big]
\end{eqnarray*}
(with implicit tensor products and implicit superscripts), where the coefficients $a_0,a_1,a_2,a_3$, and $a_4$ are, respectively, real roots of the polynomials
\begin{gather*}
 4\,608 \,x^4-1\,575 \,x^3+525 \,x^2-117 \,x-1, \\
 221\,184 \,x^4+142\,479 \,x^3-19\,701 \,x^2-15\,603 \,x+2\,363, \\
 9\,216 \,x^4-16\,857 \,x^3+11\,724 \,x^2-3\,660 \,x+430, \\
 221\,184 \,x^4-50\,895 \,x^3-16\,200 \,x^2+1\,368 \,x+602, \\
 221\,184 \,x^4+16\,335 \,x^3-37\,008 \,x^2-11400 \,x+3\,440, 
\end{gather*}
and the primed coefficients $a_0',a_1',a_2',a_3'$, and $a_4'$ are, respectively, real roots of the polynomials
\begin{gather*}
4\,608 \,x^4+8\,595 \,x^3+5\,583 \,x^2+873 \,x-43, \\
221\,184 \,x^4-101\,601 \,x^3-1\,701 \,x^2+2\,745 \,x+305, \\
3\,072 \,x^4-2\,229 \,x^3+540 \,x^2-60 \,x+4, \\
221\,184 \,x^4-294\,975 \,x^3+145\,080 \,x^2-31\,224 \,x+2\,492, \\
221\,184 \,x^4+16\,335 \,x^3-37\,008 \,x^2-11\,400 \,x+3\,440.
\end{gather*}
Numerically, their values are
\begin{eqnarray}
a_0 = 0.2744, && \qquad a_0' = 0.0390, \nonumber \\
a_1 = 0.2178, && \qquad a_1' = 0.3355, \nonumber \\
a_2 = 0.3628, && \qquad a_2' = 0.2451, \nonumber \\
a_3 = 0.3114, && \qquad a_3' = 0.4291, \nonumber \\
a_4 = a_4' = 0.2097. \hspace{-18mm} \nonumber
\end{eqnarray}

Using the same instruments for Alice and Bob as those defined in Eqs.~(\ref{def:M00}--\ref{def:M11}), our maximal probability $p_{\text{GYNI}}^{\text{max}, d=2}$ of winning the GYNI game with qubits is then found to be the smallest real root of the polynomial 
\[1\,769\,472 \,x^4 - 2\,884\,032 \,x^3 + 1\,630\,800 \,x^2 - 380\,052 \,x + 34\,087,\]
and our maximal probability $p_{\text{LGYNI}}^{\text{max}, d=2}$ of winning the LGYNI game with qubits is $p_{\text{LGYNI}}^{\text{max}, d=2} = p_{\text{GYNI}}^{\text{max}, d=2} + 1/4$. Numerically, we obtain
    \begin{equation*}
p_{\text{GYNI}}^{\text{max}, d=2} \approx 0.5694 > \frac12 \, , \quad
p_{\text{LGYNI}}^{\text{max}, d=2} \approx 0.8194 > \frac34 \, . 
    \end{equation*}

It is somewhat surprising that these maximal violations of the GYNI and LGYNI inequalities with qubits have such complicated expressions---contrary for instance to the case of Oreshkov \textit{et al.}'s original causal inequality (for which the violation exhibited in Ref.~\cite{oreshkov12} was proven, under certain constraints, to be optimal~\cite{brukner14}), or to the case of the well known CHSH Bell inequality~\cite{chsh69,tsirelson80}.
Note also that, as mentioned in the main text, we could find higher violations of the GYNI inequality using higher-dimensional quantum systems (see Table~\ref{tab:violations}), while we couldn't find any higher violations of the LGYNI inequality.

\section{\\ The set of process matrix correlations is convex}\label{app:convex}

In this Appendix we show that the set of process matrix correlations is convex.

Let $p_0(a,b|x,y)$ and $p_1(a,b|x,y)$ be two correlations realized by the (valid) process matrices and instruments $\{W_0, M_{0;\,a|x}^{A_IA_O}, M_{0;\,b|y}^{B_IB_O}\}$ and $\{W_1, M_{1;\,a|x}^{A_IA_O}, M_{1;\,b|y}^{B_IB_O}\}$, respectively, so that
\begin{eqnarray*}
p_0(a,b|x,y) &=& \tr \big[ (M_{0;\,a|x}^{A_IA_O} \otimes M_{0;\,b|y}^{B_IB_O}) \cdot W_0 \big] \,, \\
p_1(a,b|x,y) &=& \tr \big[ (M_{1;\,a|x}^{A_IA_O} \otimes M_{1;\,b|y}^{B_IB_O}) \cdot W_1 \big] \,.
\end{eqnarray*}
Without loss of generality we assume that Alice and Bob's input and output systems in $\{W_0, M_{0;\,a|x}^{A_IA_O}, M_{0;\,b|y}^{B_IB_O}\}$ and in $\{W_1, M_{1;\,a|x}^{A_IA_O}, M_{1;\,b|y}^{B_IB_O}\}$ have the same dimensions (one can indeed always embed lower-dimensional systems into larger-dimensional ones). Let us now introduce some ancillary 2-dimensional input systems with Hilbert spaces $\mathcal H^{A_I'}$ and $\mathcal H^{B_I'}$ and define, for any $q \in [0,1]$,
\begin{eqnarray*}
W &\,=\,& q \ \proj{0}^{A_I'} \otimes \proj{0}^{B_I'} \otimes W_0 \\
&& + \, (1{-}q) \ \proj{1}^{A_I'} \otimes \proj{1}^{B_I'} \otimes W_1 \,,
\end{eqnarray*}
\begin{eqnarray*}
M_{a|x}^{A_I'A_IA_O} &=& \proj{0}^{A_I'} \otimes M_{0;\,a|x}^{A_IA_O} + \proj{1}^{A_I'} \otimes M_{1;\,a|x}^{A_IA_O} \,, \\[2mm]
M_{b|y}^{B_I'B_IB_O} &=& \proj{0}^{B_I'} \otimes M_{0;\,b|y}^{B_IB_O} + \proj{1}^{B_I'} \otimes M_{1;\,b|y}^{B_IB_O} \,.
\end{eqnarray*}
One can easily check that $W$ thus defined is a valid process matrix, and that $M_{a|x}^{A_I'A_IA_O}$ and $M_{b|y}^{B_I'B_IB_O}$ define valid instruments. Furthermore, a straightforward calculation shows that for these process matrix and instruments,
\begin{eqnarray*}
p(a,b|x,y) &=& \tr \big[ (M_{a|x}^{A_I'A_IA_O} \otimes M_{b|y}^{B_I'B_IB_O}) \cdot W \big] \\
&=& q \, \tr \big[ (M_{0;\,a|x}^{A_IA_O} \otimes M_{0;\,b|y}^{B_IB_O}) \cdot W_0 \big] \\
&& + \, (1{-}q) \, \tr \big[ (M_{1;\,a|x}^{A_IA_O} \otimes M_{1;\,b|y}^{B_IB_O}) \cdot W_1 \big] \\
&=& q \, p_0(a,b|x,y) + (1{-}q) \, p_1(a,b|x,y) \,.
\end{eqnarray*}
Thus, any convex combination $q \, p_0 + (1{-}q) \, p_1$ of process matrix correlations can also be realized with a process matrix and suitable instruments, which shows that the set of process matrix correlations is indeed convex. It is straightforward to generalize the proof to a scenario with more parties.

\medskip

Note that our construction only shows that the convex hull of the set of correlations produced by process matrices of dimensions $d_{A_I} \times d_{A_O} \times d_{B_I} \times d_{B_O}$ is contained in the set of correlations produced by process matrices of dimensions $2 d_{A_I} \times d_{A_O} \times 2 d_{B_I} \times d_{B_O}$, opening up the possibility that the set of process matrix correlations is not convex for any fixed input dimension (see Figure~\ref{fig:boundary}), analogously to the set of (nonsignaling) quantum correlations for fixed dimensions~\cite{pal09}.

\bibliographystyle{linksen}
\bibliography{biblio}

\end{document}